\documentclass[twocolumn,showpacs,preprintnumbers,amsmath,amssymb]{revtex4}
\usepackage{graphicx}% Include figure files
\usepackage{dcolumn}% Align table columns on decimal point
\usepackage{bm}% bold math

\begin{document}

\preprint{}

\title{Microscopic study of spin-orbit-induced Mott insulator in Ir oxides}% Force line breaks with \\

\author{Hiroshi Watanabe$^{1,2}$}
 \email{h-watanabe@riken.jp}
\author{Tomonori Shirakawa$^{1,2}$}
\author{Seiji Yunoki$^{1,2,3}$}
\affiliation{%
$^1$Computational Condensed Matter Physics Laboratory, RIKEN ASI, Wako, Saitama 351-0198, Japan\\
$^2$CREST, Japan Science and Technology Agency (JST), Kawaguchi, Saitama 332-0012, Japan\\
$^3$Quantum Systems Materials Science Research Team, RIKEN AICS, Kobe, Hyogo 650-0047, Japan
}%

\date{\today}% It is always \today, today,

\begin{abstract}
Motivated by recent experiments of a novel 5$d$ Mott insulator in Sr$_2$IrO$_4$, 
we have studied the two-dimensional three-orbital Hubbard model with a spin-orbit coupling $\lambda$.  
The variational Monte Carlo method is used to obtain the ground state phase diagram with varying a on-site 
Coulomb interaction $U$ as well as $\lambda$. It is found that the transition from a paramagnetic metal 
to an antiferromagnetic (AF) insulator occurs at a finite $U=U_{\mathrm{MI}}$, which is greatly reduced 
by a large $\lambda$, characteristic of 5$d$ electrons, and leads to the ``spin-orbit-induced'' Mott insulator. 
It is also found that the Hund's coupling induces the anisotropic spin exchange and stabilizes the in-plane AF order. 
We have further studied the one-particle excitations using the variational cluster approximation, and revealed 
the internal electronic structure of this novel Mott insulator. 
These findings are in agreement with experimental observations on Sr$_2$IrO$_4$, and qualitatively 
different from those of extensively studied 3$d$ Mott insulators. 

\end{abstract}

\pacs{71.30.+h, 75.25.Dk, 71.20.-b}% PACS, the Physics and Astronomy
\maketitle

Transition metal oxides have been one of the most fascinating classes of materials in recent 
years~\cite{Imada}. 
For the past decays, tremendous amount of efforts have been devoted to explore the nature 
of 3$d$ transition metal oxides where various exotic states and phenomena have emerged such as 
high-$T_{\rm c}$ cuprate superconductors, colossal magneto-resistant manganites, multiferroics,  
and various magnetic orders. It has been established that these states and phenomena are caused by 
strong Coulomb interactions along with cooperative interactions of spin, charge, and orbital degrees 
of freedom, which are basically separable in 3$d$ electrons~\cite{Tokura}. 

Very recently, 5$d$ transition metal oxides have attracted much attention as a candidate of a novel 
Mott insulator. Because of the extended nature of 5$d$ orbital, Coulomb interactions are expected to be 
smaller for 5$d$ electrons ($\sim$ 1--3 eV) than for 3$d$ electrons ($\sim$ 5--7 eV)~\cite{Taguchi}, whereas the spin-orbit 
coupling (SOC) $\lambda$ is estimated to be considerably larger in 5$d$ ($\sim$ 0.1--1 eV) than in 
3$d$ ($\sim$ 0.01--0.1 eV). Therefore, in 5$d$ transition metal oxides, inherently 
spin and orbital degrees of freedom are strongly entangled.  

One of such examples is the layered 5$d$ transition metal oxide Sr$_2$IrO$_4$.  
In Sr$_2$IrO$_4$, $t_{2g}$ and $e_g$ orbitals are well separated by the large crystal field, 
and the lower $t_{2g}$ orbital is filled with five electrons, $(t_{2g})^5$. 
In spite of the large band width and small on-site Coulomb interaction $U$, Sr$_2$IrO$_4$ is an antiferromagnetic 
insulator with a weak ferromagnetic moment~\cite{Crawford,Cao}.  
Neutron diffraction patterns do not detect any superlattice structure that indicates charge order or 
charge density wave states~\cite{Huang}. It is proposed that the strong SOC is responsible for the insulating 
mechanism~\cite{Kim1}. 
Indeed, the 4$d$ counterpart of Sr$_2$RhO$_4$, which has a larger $U$ and a smaller $\lambda$ than 
Sr$_2$IrO$_4$, is metallic~\cite{Baumberger}. 

\begin{figure}
\begin{center}
\includegraphics[width=0.9\hsize]{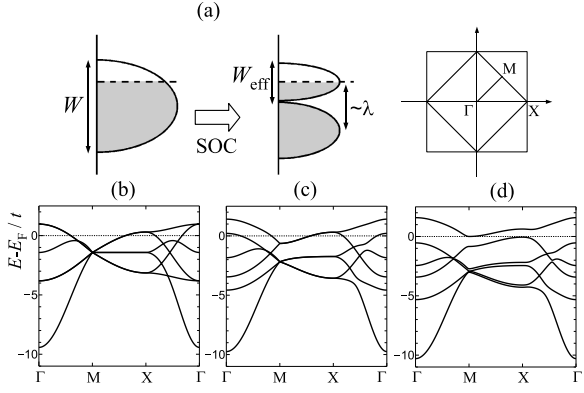}
\caption{\label{fig1}(a) Schematic picture of the splitting of density of states by the SOC.
Non-interacting tight-binding energy band (b) without the SOC, (c) with the SOC 
($\lambda/t=1.028$), and (d) with the SOC and staggered AF field. 
See the text for the tight-binding parameters used.} 
\end{center}
\end{figure}

The proposed picture of this ``spin-orbit-induced'' Mott insulator in Sr$_2$IrO$_4$ 
is as follows. 
The SOC splits the $t_{2g}$ orbitals into $J_{\mathrm{eff}}=1/2$ states ($J_{\mathrm{eff}}^z=\pm1/2$, twofold degenerate) 
and $J_{\mathrm{eff}}=3/2$ states ($J_{\mathrm{eff}}^z=\pm1/2,\pm3/2$, fourfold degenerate).  
Here $J_{\mathrm{eff}}$ denotes the ``effective'' total angular momentum derived from the large SOC 
with the large crystal field~\cite{Kim1}.  
When the SOC is large enough, the lower $J_{\mathrm{eff}}=3/2$ state is fully filled and 
the upper $J_{\mathrm{eff}}=1/2$ state forms an active half-filled energy band. 
The band width of this half-filled band ($W_{\mathrm{eff}}$) is much smaller than 
the original one without the SOC ($W$) as shown in Fig.~\ref{fig1}(a), and thus even small $U$ can lead the system into a Mott insulator.  
This picture of ``$J_{\mathrm{eff}}=1/2$ Mott insulator'' is supported by several experiments such as resonant x-ray scattering~\cite{Kim2},  
angle resolved photoemission spectroscopy~\cite{Kim1}, and optical conductivity~\cite{Kim1}.  
This novel Mott insulator has a quite different character than 3$d$ Mott insulators where the effect of the SOC is only 
perturbative and thus the spin and orbital are essentially separate objects. 
Although this picture is supported by first-principle calculations based on the density functional theory~\cite{Kim1,Jin}, 
a study treating many-body effects beyond the mean-filed level is still lacking. The main purpose of this paper is 
to understand the nature of Mott insulator induced in spin-orbital entangled 5$d$ systems. 

Here, we study the novel Mott transition and magnetic order induced by the SOC for Sr$_2$IrO$_4$. 
The ground state properties of the three-orbital Hubbard model with a SOC term are investigated with the variational 
Monte Carlo (VMC) method. We show that the large SOC works cooperatively with $U$ and leads the system into a novel 
Mott insulating state with an in-plane AF order. 
We also calculate the one-particle excitation spectrum using the variational cluster approximation (VCA)~\cite{vca} 
to discuss 
the difference between the ``spin-orbit-induced'' Mott insulator and the 3$d$ Mott insulator.

We take the three-orbital Hubbard model on a two-dimensional (2D) square lattice defined by the following Hamiltonian 
$H=H_{\mathrm{kin}}+H_{\mathrm{SO}}+H_{\mathrm{I}}$, where 
$H_{\mathrm{kin}}=\sum_{\bm{k}\alpha\sigma}\varepsilon_{\alpha}(\bm{k})c_{\bm{k}\alpha\sigma}^{\dagger}
        c_{\bm{k}\alpha\sigma}$ is the kinetic term, 
$H_{\mathrm{SO}}=\lambda\sum_{i}\bm{L}_i\cdot\bm{S}_i$ is the SOC term with a coupling constant $\lambda$, and 
\begin{align}
H_{\mathrm{I}}&=U\sum_{i,\alpha}n_{i\alpha\uparrow}n_{i\alpha\downarrow}, \notag \\
     &+\sum_{i,\alpha<\beta,\sigma}\left[U'n_{i\alpha\sigma}n_{i\beta-\sigma}
  +(U'-J)n_{i\alpha\sigma}n_{i\beta\sigma}\right] \notag \\
 &+J\sum_{i,\alpha<\beta,\sigma}(c^{\dagger}_{i\alpha\uparrow}c^{\dagger}_{i\beta\downarrow}
  c_{i\alpha\downarrow}c_{i\beta\uparrow}
  +c^{\dagger}_{i\alpha\uparrow}c^{\dagger}_{i\alpha\downarrow}
   c_{i\beta\downarrow}c_{i\beta\uparrow}+\mathrm{H.c.}) %\nonumber
   \label{int}
\end{align}
is the Coulomb interactions including intraorbital, interorbital, and spin-flip and pair-hopping interactions~\cite{kanamori}. 
Here, $\sigma$ indicates electronic spins and the indices $\alpha$ and $\beta$ 
represent three $t_{2g}$ orbitals, $yz$ (1), $zx$ (2), and $xy$ (3). 

The kinetic and SOC terms can be combined ($H_0=H_{\mathrm{kin}}+H_{\mathrm{SO}}$) in the matrix form,
\begin{multline}
H_0=\sum_{\bm{k}\sigma}\left(c^{\dagger}_{\bm{k}1\sigma},c^{\dagger}_{\bm{k}2\sigma},c^{\dagger}_{\bm{k}3-\sigma}\right) \\
\times
\begin{pmatrix}
\varepsilon_1(\bm{k}) & \mathrm{i}\sigma\lambda/2 & -\sigma\lambda/2 \\
-\mathrm{i}\sigma\lambda/2 & \varepsilon_2(\bm{k}) & \mathrm{i}\lambda/2 \\
-\sigma\lambda/2 & -\mathrm{i}\lambda/2 & \varepsilon_3(\bm{k})
\end{pmatrix}
\begin{pmatrix}
c_{\bm{k}1\sigma} \\
c_{\bm{k}2\sigma} \\
c_{\bm{k}3-\sigma}
\end{pmatrix},
\end{multline}
from which it is apparent that the SOC mixes the up- and down-spin states, and new quasiparticles are obtained 
simply by diagonalizing $H_0$. The new quasiparticles are characterized by the pseudo-orbital $\alpha$ and 
pseudo-spin $\sigma$ with a creation (annihilation) operator $a^{\dagger}_{\bm{k}\alpha\sigma}$ 
($a_{\bm{k}\alpha\sigma}$). In the atomic limit ($\varepsilon_1(\bm{k})=\varepsilon_2(\bm{k})=\varepsilon_3(\bm{k})=0$), 
sixfold degenerate states are split into twofold degenerate $J_{\mathrm{eff}}=1/2$ states and fourfold degenerate $J_{\mathrm{eff}}=3/2$ 
states~\cite{Kim1}. 

First, we construct the non-interacting tight-binding energy band.
In Sr$_2$IrO$_4$, and also in Sr$_2$RhO$_4$, there is a large hybridization between the $xy$ and $x^2-y^2$ orbitals 
originated from the tilting of IrO$_6$ octahedra.
Because of this hybridization, the $xy$ orbital is pushed down below the Fermi energy~\cite{Ru}.
To take into account the effect of this hybridization, we introduce the next-nearest and third-nearest hopping integrals 
in the $xy$ orbital. The chemical potential $\mu_3$ is also introduced to take account of both the hybridization 
and tetragonal splitting~\cite{Jin}. 
The resulting energy dispersion of the $xy$ orbital is 
$\varepsilon_3(\bm{k})=-2t_1(\cos k_x+\cos k_y)-4t_2\cos k_x\cos k_y-2t_3(\cos 2k_x+\cos 2k_y)+\mu_3$. On the contrary, 
the $yz$ and $zx$ orbitals have almost one-dimensional character, $\varepsilon_1(\bm{k})=-2t_5\cos k_x-2t_4\cos k_y$ and
$\varepsilon_2(\bm{k})=-2t_4\cos k_x-2t_5\cos k_y$ ($t_4\gg t_5$). 
Using this form, the band dispersion of Sr$_2$IrO$_4$ calculated by the LDA+SO (spin-orbit)~\cite{Kim1} is 
well reproduced by choosing the tight-binding parameters with 
($t_1, t_2, t_3, t_4, t_5, \mu_3, \lambda$)=(0.36, 0.18, 0.09, 0.37, 0.06, -0.36, 0.37) eV, 
as shown in Fig.~\ref{fig1}(c).  
In the following, we set $t_1=t$ as an energy unit.
To study the Mott transition and the role of the SOC, we only change the value of $\lambda$ and fix the other tight-binding parameters. 
This assumption is justified by the fact that the LDA+SO band dispersion for the 4$d$ counterpart of metallic 
Sr$_2$RhO$_4$~\cite{Liu} is well reproduced by choosing $\lambda /t\sim0.5$ with the other parameters fixed. 

Next, we examine the effect of Coulomb interactions. For this purpose, the following trial wave function is considered,
\begin{equation}
\left|\Psi \right>=P_{\mathrm{J_c}}P_{\mathrm{G}}\left|\Phi \right>.
\label{twf}
\end{equation}
$\left|\Phi \right>$ is the one-body part obtained by diagonalizing $H_0$ with renormalized (variational) tight-binding parameters, 
${\bar H}_0(\tilde{t}_i, \tilde{\mu}_3, \tilde{\lambda}_{\alpha\beta})$.
Note that by the effect of Coulomb interaction and tetragonal splitting, the ``effective'' coupling constant of the SOC 
has orbital dependence: $\lambda\rightarrow\tilde{\lambda}_{\alpha\beta}$.
To consider magnetically ordered states, a term with the different magnetic order parameter is added to ${\bar H}_0$.
Here, we study the out-of-plane AF order (along $z$ axis) and in-plane AF order (along $x$ axis), described by 
$\Delta^z\sum \mathrm{e}^{\mathrm{i}\bm{Q}\cdot\bm{r}_i}(a^{\dagger}_{i\alpha\uparrow}a_{i\alpha\uparrow}
-a^{\dagger}_{i\alpha\downarrow}a_{i\alpha\downarrow})$ and 
$\Delta^x\sum \mathrm{e}^{\mathrm{i}\bm{Q}\cdot\bm{r}_i}(a^{\dagger}_{i\alpha\uparrow}a_{i\alpha\downarrow}
+a^{\dagger}_{i\alpha\downarrow}a_{i\alpha\uparrow})$, respectively. Here $\bm{Q}=(\pi,\pi)$ is an ordering vector. 
Note that the staggered field is applied to newly formed quasiparticles in the real space ($a_{i\alpha\sigma}, a^{\dagger}_{i\alpha\sigma}$)
and $\sigma$ represents the pseudo-spin, not the original spin.
The matrix to be diagonalized at each $\bm{k}$ becomes 12$\times$12 for the AF state (not shown). 

The operator $P_{\mathrm{G}}=\prod_{i,\gamma}\left[1-(1-g_\gamma)\left|\gamma\right>\left<\gamma\right|_i\right]$ 
in Eq.~(\ref{twf}) is a Gutzwiller factor extended for the three-orbital system~\cite{Bunemann}.
Here, $i$ is a site index and $\gamma$ runs over possible electron configurations at each site.  For the three-orbital system, 
there are $4^3=64$ electron configurations, namely, $\left|0\right>=\left|0\;0\;0\right>$,
$\left|1\right>=\left|0\;0\uparrow\right>$, $\cdots$, 
$\left|63\right>=\left|\uparrow\downarrow\;\uparrow\downarrow\;\uparrow\downarrow\right>$.
We control the weight of each electron configuration by varying $g_{\gamma}$ from 0 to 1. 
The set of $\{{g_{\gamma}}\}$ is a variational parameter and optimized so as to give the lowest energy. 
In this study, we classify the possible 64 patterns into 12 groups by the Coulomb interaction energy, and $g_\gamma$'s in the same group 
are set to be the same. 

The remaining term $P_{\mathrm{J_c}}=\exp\left[-\sum_{i\neq j}v_{ij}n_in_j\right]$ in Eq.~(\ref{twf}) is a charge Jastrow 
factor that controls the long-range charge correlation. 
The long-range charge correlation is known to be important for describing Mott transition~\cite{Capello}. 
In this study, we assume that $v_{ij}$ depends only on the 
distances, $v_{ij}=v(|\bm{r}_i-\bm{r}_j|)$. 
%Since $i$ and $j$ run over all lattice sites, the number of variational parameters increases with the system size.
For instance, we consider up to 19th-neighbor correlation for a 10$\times$10 square lattice, and therefore the number of 
independent variational parameters $v_{ij}$ is 19.

The ground state energy and other physical quantities are calculated with the VMC method.
The variational parameters mentioned above are simultaneously optimized by using the stochastic reconfiguration method~\cite{Sorella}. 
This method makes it possible to optimize many parameters efficiently and stably.
%Once the trial wave function is determined, we can easily calculate physical quantities.  

\begin{figure}
\includegraphics[width=\hsize]{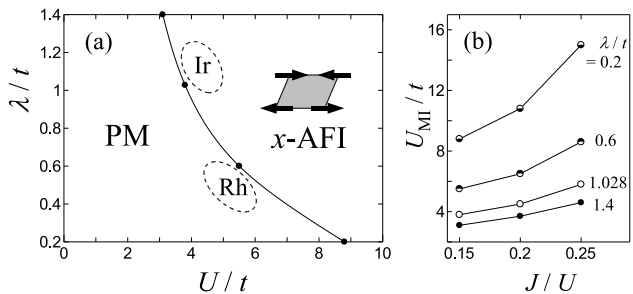}
\caption{\label{fig2}(a) Ground state phase diagram of the three-orbital Hubbard model in a 2D 10$\times$10 square lattice 
with $n=5$, $U'/U=0.7$, and $J/U=0.15$. 
PM denotes the paramagnetic metal and $x$-AFI denotes the AF insulator with an in-plane (along $x$ axis) magnetic moment.
The solid line indicates the first-order phase boundary $U_{\mathrm{MI}}$. 
The expected locations of Sr$_2$IrO$_4$ (Ir) and Sr$_2$RhO$_4$ (Rh) are also indicated in the phase diagram. 
(b) $J/U$ dependence of $U_{\mathrm{MI}}$ for different $\lambda$.}
\end{figure}

Fig.~\ref{fig2}(a) shows the obtained ground state phase diagram for a 2D 10$\times$10 square lattice 
with the electron density $n=5$, corresponding to $(t_{2g})^5$, and the Coulomb interactions $U'/U=0.7$ and $J/U=0.15$. 
The transition from the paramagnetic metal to the AF insulator occurs at $U=U_{\mathrm{MI}}$, which depends sensitively on the value 
of $\lambda$. This metal-insulator transition (MIT) is found to be first-order, indicating that the nesting of 
the Fermi surface is not perfect and the Coulomb interaction is essential in driving the transition. 
The insulating mechanism is understood as follows. 
When the staggered (AF) field is applied, the degeneracy along the edge of the AF Brillouin zone
(along M-X in Fig.~\ref{fig1}) is lifted and the highest band is split off by the AF gap [Fig.~\ref{fig1}(d)].
As can be seen in Figs.~\ref{fig1}(b) and \ref{fig1}(c), the SOC lifts upward the two branches of the energy bands 
from the rest, which makes it easier to fully open the AF gap once the Coulomb interactions are considered. 
Namely, the larger the SOC is, the easier the system becomes the AF insulator. 
Indeed, Fig.~\ref{fig2}(a) shows that $U_{\mathrm{MI}}$ monotonically decreases with increasing $\lambda$. 
Moreover, we found that the effective coupling constant $\tilde{\lambda}_{\alpha\beta}$ increases with $U$, 
indicating that $U$ and $\lambda$ work cooperatively for insulating. 

We expect that this insulating mechanism found above is applied for Sr$_2$IrO$_4$. 
Because the Coulomb interaction $U_{\mathrm{Ir}}$ is much smaller than the band width, the metallic state is naively expected 
for Sr$_2$IrO$_4$. However, Sr$_2$IrO$_4$ is experimentally found insulating with a canted AF order~\cite{Kim2}. 
This counter-intuitive observation can be explained if the SOC in Sr$_2$IrO$_4$, $\lambda_{\mathrm{Ir}}$, is large enough to 
reduce $U_{\mathrm{MI}}$ smaller than $U_{\mathrm{Ir}}$. 
Indeed, $\lambda_{\mathrm{Ir}}$ is estimated as large as 0.4--0.5 eV, which is much larger than the values for 
3$d$ and 4$d$ electron systems.
We consider that $U_{\mathrm{Ir}}>U_{\mathrm{MI}}$ is satisfied and the ``spin-orbit-induced" Mott insulator is 
realized in Sr$_2$IrO$_4$. On the other hand, in Sr$_2$RhO$_4$, we consider that $U_{\mathrm{Rh}}<U_{\mathrm{MI}}$ is 
satisfied and thus the metallic state is realized. The expected locations of Sr$_2$IrO$_4$ and Sr$_2$RhO$_4$ in the phase 
diagram are indicated in Fig.~\ref{fig2}(a), where both of them are located near the MIT point. 
Note that the observed insulating gap of Sr$_2$IrO$_4$ is very small ($\sim$ 0.1 eV), suggesting that the system is in the vicinity 
of the MIT point. It is also interesting to notice that a small amount of substitution of Ir ($\sim$ 10$\%$) for Rh in Sr$_2$RhO$_4$ 
leads to the MIT~\cite{Yoshida}, indicating that the system is located not far from the MIT point. 

The $J/U$ dependence of $U_{\mathrm{MI}}$ is also investigated and the results are shown in 
Fig.~\ref{fig2}(b). It is clearly seen in Fig.~\ref{fig2}(b) that $U_{\mathrm{MI}}$ monotonically increases with increasing $J/U$, 
indicating that the Hund's coupling is unfavorable for the spin-orbit-induced Mott insulator.
This behavior is naturally understood since the Hund's coupling competes with the SOC and works destructively for the formation 
of the quasiparticles originated from the SOC. It is, however, found that the overall shape of the phase diagram does not change 
qualitatively with increasing $J/U$ except for $U_{\mathrm{MI}}$ shifting to a larger value. 
For $J/U=$0.15--0.25, we estimate $U_{\mathrm{MI}}=$1.2--1.6 eV for Sr$_2$IrO$_4$ and 1.6--2.4 eV for Sr$_2$RhO$_4$. 

In the insulating region, we found that the in-plane AF order ($x$-AFI) is more stable than the out-of-plane AF order ($z$-AFI). 
If there is no Hund's coupling [$J=0$ in Eq.~(\ref{int})], the rotational symmetry in pseudo-spin space is preserved and 
$z$-AFI and $x$-AFI are energetically degenerate.  
However, the introduction of Hund's coupling induces the anisotropy and the in-plane AF order is more favored 
than the out-of-plane AF order. 
This result is consistent with the study of an effective strong SOC spin model by Jackeli and Khaliullin~\cite{Jackeli}. 
The magnetic x-ray diffraction experiment~\cite{Kim2} also supports the in-plane magnetism in Sr$_2$IrO$_4$. 

We have also estimated the local magnetic moment as large as $0.3$--$0.4\mu_{\mathrm{B}}$ for the parameters appropriate 
for Sr$_2$IrO$_4$. This value is comparable to the results of magnetic susceptibility measurements~\cite{Cao,Kini} and 
much smaller than the atomic value of 1 $\mu_{\mathrm{B}}$ in the strong-SOC limit. 
This large reduction is due to the large itinerancy of 5$d$ electrons. %and the system is located far from the fully-ordered AF state. 
If we assume that the magnetic moment exactly follows the tilting of IrO$_6$ octahedra observed experimentally 
($\sim$ 11$^{\circ}$)~\cite{Crawford}, the expected ferromagnetic moment is 0.05--0.07$\mu_{\mathrm{B}}$, which is comparable to 
the experimental estimation~\cite{Crawford,Cao,Kini}. 

\begin{figure}[t]
\includegraphics[width=0.8\hsize]{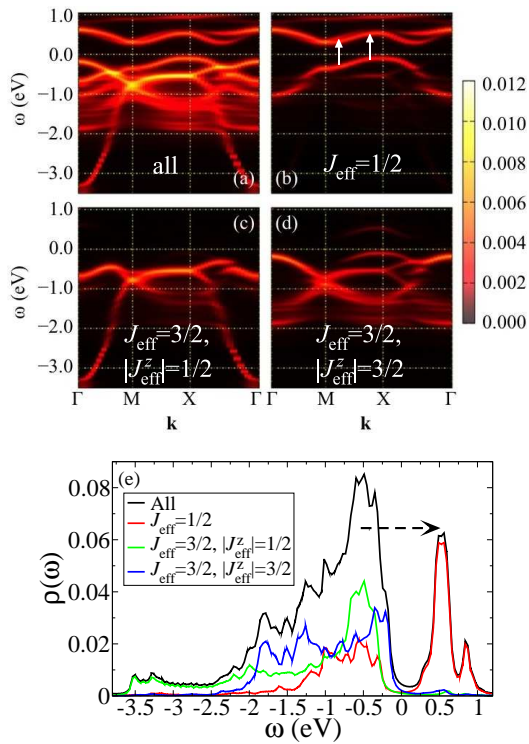}
\includegraphics[width=0.7\hsize]{fig3-2.eps}
\caption{\label{fig3}(color online). (a) One-particle excitation spectrum for the Mott insulating state.
Spectra projected onto (b) $J_{\mathrm{eff}}=1/2$, 
(c) $J_{\mathrm{eff}}=3/2, J_{\mathrm{eff}}^z=\pm1/2$, and (d) $J_{\mathrm{eff}}=3/2, J_{\mathrm{eff}}^z=\pm3/2$ states as well as 
(e) density of states $\rho(\omega)$ are also shown. The parameters used are $U$=1.44 eV, $U'$=1.008 eV, $J$=0.216 eV, and 
$\lambda$=0.432 eV, and $\omega=0$ corresponds to the Fermi energy. }
\end{figure}

Finally, to explore the internal electronic structure in the insulating state, the one-particle excitation spectrum is calculated using 
the VCA~\cite{note}. 
The results are shown in Fig.~\ref{fig3} for a set of parameters appropriate for Sr$_2$IrO$_4$. 
As seen in Fig.~\ref{fig3}(a), the Fermi energy is located inside the gap, and thus the state is insulating. 
The validity of the physical picture of ``$J_{\mathrm{eff}}=1/2$ Mott insulator'' can be examined by projecting the spectrum onto 
$J_{\mathrm{eff}}=1/2$ [Fig.~\ref{fig3}(b)] and $J_{\mathrm{eff}}=3/2$ [Figs.~\ref{fig3}(c) and~\ref{fig3}(d)] states defined in 
the atomic limit. It is clearly seen in Fig.~\ref{fig3} that 
the upper (unoccupied) band is well described by $J_{\mathrm{eff}}=1/2$ state. 
On the other hand, the lower (occupied) band is a mixture of $J_{\mathrm{eff}}=1/2$ and $J_{\mathrm{eff}}=3/2$ states and can 
not be simply compared with the atomic limit. For example, 
the bands with $J_{\mathrm{eff}}=3/2$ character are located closer to the Fermi energy 
for all momenta except for the vicinity of the AF Brillouin zone boundary (M-X) 
[Figs.~\ref{fig3}(b) and (d)]. In this sense, the low-lying one-particle excitations are slightly different from the picture of 
``$J_{\mathrm{eff}}=1/2$ Mott insulator'' where the $J_{\mathrm{eff}}=1/2$ and 
$J_{\mathrm{eff}}=3/2$ states are well separated and the former forms the lowest-lying upper and lower Hubbard bands. 
This difference is attributed to the large itinerancy of 5$d$ electrons.

Fig.~\ref{fig3}(e) shows the total density of states calculated using the VCA. 
As indicated by the arrow in Fig.~\ref{fig3}(e), the optical excitation is expected mainly at around 1 eV originating 
from $\sim-$0.5 eV to $\sim$ 0.5 eV with a broad width due to a large band width of the occupied states. 
Indeed, a moderate peak structure is observed at around 1 eV in the optical conductivity measurement~\cite{Kim2}, which is 
in good agreement with our calculations. In addition, the experiment has observed a rather sharp peak at around 0.5 eV. 
This sharp peak may be explained by the excitation from the lower $J_{\mathrm{eff}}=1/2$ state 
to the upper $J_{\mathrm{eff}}=1/2$ state along the AF Brillouin zone boundary (M-X), indicated by the arrows in Fig.~\ref{fig3}(b), 
where the almost parallel bands expect to induce a sharp excitation ($\sim$ 0.5 eV). 

In conclusion, we have shown that the SOC plays an important role in determining sensitively the critical value of $U_{\rm MI}$ 
as well as the nature of the resulting Mott insulating phase in Sr$_2$IrO$_4$. This is because the 5$d$ Mott insulator, not like 
for 3$d$ systems, is well described by the novel quantum number $J_{\mathrm{eff}}$, which is due to the large SOC along with 
the large crystal field, a generic feature for the 5$d$ transition metal oxides. Therefore, we expect that not only $U$ but also 
$\lambda$ are crucial factors to be considered in describing MIT for the 5$d$ systems in general. 
Moreover, because of the orders of magnitude different $\lambda$, 
other novel quantum phenomena, such as the anomalous metallic state, 
multipole order, and unconventional superconductivity that are not observed in the 3$d$ systems, are 
expected to emerge in the 5$d$ systems.  

%\begin{acknowledgments}
The authors thank Y. Yanase, M. Taguchi, and A. Rusydi for useful discussions.
A part of this work is supported by CREST (JST).
The computation in this work has been done using the RIKEN Cluster of Clusters (RICC) facility and the facilities of the Supercomputer Center,
Institute for Solid State Physics, University of Tokyo.
%\end{acknowledgments}


\begin{thebibliography}{99}
\bibitem{Ru} It is a quite different band structure from the superconducting counterpart of Sr$_2$RuO$_4$, 
where the $xy$-originated $\gamma$ band greatly contributes the superconducting behavior.
\bibitem{note} A $\sqrt{2}\times\sqrt{2}$ cluster is used. However, the results discussed here do not change qualitatively for different sizes of clusters. 
We have confirmed that the Mott transition is properly described by the VCA. The fact that the two different methods (VMC and VCA) used 
give the same results indicates the validity of the spin orbit-induced Mott insulator.
\end{thebibliography}
\end{document}